\documentclass{article}
\usepackage[utf8]{inputenc}

\usepackage{geometry}
\usepackage{wrapfig}
\usepackage{multicol,caption}
\usepackage{titlesec}
\usepackage[T1]{fontenc}
\usepackage{times}
\usepackage{setspace}

\usepackage{amsmath}
\usepackage{latexsym}
\usepackage{amsfonts}
\usepackage{mathrsfs}
\usepackage{upgreek}
\usepackage{graphics}
\usepackage{graphicx}
\usepackage{epstopdf}
\usepackage{epsfig}
\usepackage{amsbsy}
\usepackage{amsfonts}
\usepackage{amsmath}
\usepackage{amssymb}
\usepackage{bm}
\usepackage{color}

\title{Analysis of a constant height hydraulic fracture}

\author{E.V. Dontsov}

\begin{document}
\maketitle

\begin{abstract}
    \noindent This study analyzes the problem of a constant height hydraulic fracture. It is assumed that the fracture is driven by Newtonian fluid, and the effects of fracture toughness and leak-off are included in the model as well. Analysis of the tip region for such a fracture is presented first. The limiting solutions and their locations in the two-dimensional parametric space are investigated. In addition, an approximate solution for the problem is constructed and its accuracy is examined in the whole parametric space. Then, analysis of the finite fracture is presented. Analytic expressions for the limiting solutions are obtained. Full numerical solution is constructed, as well as an approximation, which is based on the solution for the tip region and the global volume balance, is developed. The numerical solution is used to check accuracy of the approximation, while the latter is used to investigate the whole parametric space for the problem. This in turn allows determining zones of applicability of the limiting solutions that can be further used to quantify which problem parameters dominate fracture dynamics. 
\end{abstract}
  
\section{Introduction}
  
Hydraulic fractures are most often created underground for the purpose of enhancing oil and gas recovery from reservoirs~\cite{Econo2000}. Nowadays, complex numerical models are readily used to model such a phenomenon, which can consider propagation of a single or multiple hydraulic fractures, interaction between fractures, proppant transport, thermal effects, chemical effects, multiphase flow effects, etc. Yet, understanding of basic mechanisms and behavior of hydraulic fractures is better obtained by using simple models that focus on a particular aspect of the problem, while ignore the rest of the complexity. This is the case for this study, in which a relatively simple geometry of a constant height hydraulic fracture is investigated.

There is a variety of ``simple'' hydraulic fracture models. The word ``simple'' here essentially means that the mathematical formulation for the problem is one-dimensional. As a result of such mathematical simplicity, it is possible to investigate the problem in greater details and to better understand structure of the solution. The first example is the problem of a semi-infinite hydraulic fracture, which is the model for the tip region of a finite planar fracture~\cite{Detou2003,Gara2011,Peir2008}. The limiting analytic solutions for the problem were first obtained in~\cite{Dessr1994} and~\cite{Leno1995} for the case of domination of viscosity and leak-off, respectively. The last limiting solution that corresponds to the domination of toughness stems from classical Linear Elastic Fracture Mechanics (LEFM)~\cite{Rice1968}. General solution for the problem that considers Newtonian fluid, Carter's leak-off, and toughness was obtained in~\cite{Gara2011}. The rapid approximation for this solution was later derived in~\cite{Dont2015d}, which then was subsequently used as a propagation condition in a planar hydraulic fracture model~\cite{Dont2017a}. It is also worth mentioning multiple extensions of the model that consider power-law fluids~\cite{Dont2018}, Herschel-Bulkley fluids~\cite{Bessm2019}, Carreau fluids~\cite{moukhtari2018semi}, the effect of fluid lag~\cite{Gara2000}, the effect of turbulent flow~\cite{Dont2016d}, the effect of cohesive zone~\cite{garagash2019cohesive}, the effect of proppant~\cite{bessmertnykh2020effects}, and the effect of pressure dependent leak-off~\cite{kanin2020near}. This model already shows how multiple physical effects can be understood by considering a relatively simple scenario. 

Plane strain and radial hydraulic fractures can be considered ``simple'' geometries, since the mathematical formulation for them is also one-dimensional. In contrast to the semi-infinite fracture, for which there are three limiting solutions, the finite fractures have four limiting cases. Note that this is only the case for the model that considers the effects of toughness, viscosity and Carter's leak-off, i.e. ignoring the effects of fluid lag, cohesive zone, complex fluid rheology, etc. The aforementioned four cases are defined by the competition between fracture toughness and fluid viscosity, and fluid storage inside the fracture or inside the formation. Therefore these limits are defined as: storage toughness, storage viscosity, leak-off toughness, and leak-off viscosity, see e.g. review paper~\cite{Detou2016}. Early studies focused predominantly on finding solutions for these limiting cases~\cite{Savi2001,Adachi2002,Bung2005,Adachi2008b}. The first analysis for a complete problem is given by~\cite{Hu2010} for the case of a plane strain fracture and~\cite{MadyarPhD} for the case of a radial geometry. An alternative analysis for these problems was introduced in~\cite{Dont2016f,Dont2017c}. The rapid approximate solution for the global problem is constructed by using the semi-infinite or tip solution~\cite{Dont2015d} and global volume balance. Such rapid solution allowed to scan the whole parametric space and to construct map of the solutions that determines regions of applicability of the limiting solutions. Similar concept was later used to analyze the effect of anisotropy and power-law rheology on the parametric space in~\cite{Dont2018e}, to investigate the effect of cohesive zone in~\cite{garagash2019cohesive}, and to construct an ultra-fast hydraulic fracturing simulator in~\cite{Dont2019}.

The last, but not least, ``simple'' geometry is a constant height hydraulic fracture or Perkins-Kern-Nordgren (PKN) fracture~\cite{Perk1961,Nord1972}. Technically, it is possible to also include pseudo-3D model~\cite{Sett1986,Adachi2010,Dont2015c} into the list of ``simple'' cases. But it is noticeably more complex in the sense that its parametric space is larger, which makes the analysis more sophisticated. The classical PKN model does not consider the effect of fracture toughness, only viscosity and leak-off. The viscosity and leak-off dominated solutions can be found in~\cite{Econo2000}. Analysis of the tip region, again in the absence of toughness, is done in~\cite{Koval2010}. Laboratory observations that confirm the predicted near-tip behavior are summarized in~\cite{Xing2017}. Note that there is also analysis of the effects of turbulent flow~\cite{Zolf2017,Zia2017} for this fracture geometry. All of the above studies do not consider the effect of fracture toughness on the solution. There are two possibilities to include toughness into the model. The first was proposed by~\cite{Nolte1991}, in which the pressure boundary condition at the tip is taken from that for a uniformly pressurized radial fracture. This condition was later revised in~\cite{Sarva2015}, where authors proposed to alter it slightly based on energetic considerations. The second approach is to replace the local elasticity relation used in the classical PKN model by a more accurate non-local expression, that can be combined with the standard LEFM propagation criterion~\cite{Dont2016a}. While the second approach is shown to be more accurate~\cite{Dont2016a}, it is more difficult to do analysis with it. Therefore, the model with local elasticity and a finite pressure boundary condition at the tip is used in this study.

As is evident from the above literature review, the parametric spaces for all ``simple'' models except the PKN case were thoroughly analyzed for the case when the model considers the simultaneous interplay between toughness, viscosity, and leak-off. This is probably because the procedure to include the effect of toughness into PKN model was introduced relatively recently. To fill the gap, this study aims to investigate the parametric space for the tip region of a PKN fracture, i.e. extend the work of~\cite{Koval2010} to include the effect of fracture toughness. In addition to that, the bigger goal is to thoroughly investigate the parametric space for the finite fracture and to determine all the limiting solutions and their applicability zones in the parametric space. In particular, the structure of this study is the following. Section~\ref{goveq} outlines the governing equations for the problem. Then, the problem of the tip region is analyzed in section~\ref{sectiontip} and the parametric map for the problem is constructed. After that, section~\ref{sectionvertex} presents vertex solutions for a finite PKN fracture. The parametric map for the finite fracture is constructed in section~\ref{sectionfull} using the fast approximate solution that is benchmarked against the numerical solution. Finally, section~\ref{sectionapplex} presents application examples and section~\ref{sectionsumm} summarizes the results.

\section{Governing equations}\label{goveq}

\begin{figure}[h]
\centering\includegraphics[width=0.95\linewidth]{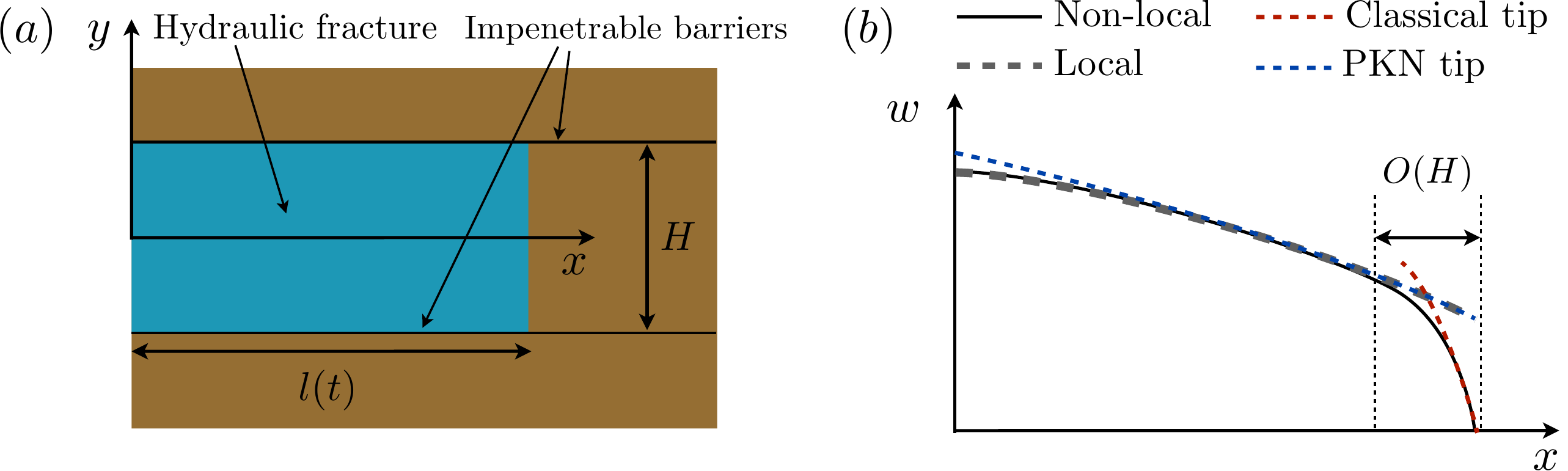}
\caption{Schematics of a constant height hydraulic fracture (panel $(a)$). Schematics of the variation of fracture width along the $x$ axis for different models (panel $(b)$).}
\label{figschem}
\end{figure}

Consider the problem of a constant height hydraulic fracture or simply PKN fracture depicted in Fig.~\ref{figschem}$(a)$. Let the fracture be vertical and occupy the $(x,y)$ plane, whereby $y$ is the vertical coordinate, while $x$ is the horizontal coordinate. Let $H$ be the fracture height and $l$ be the half-length. The height $H$ is assumed to be constant for the whole fracture, while the length is a function of time. It is also assumed that $H\ll l$, which allows to simplify the governing equations to be one-dimensional.

Various solutions for the problem and their asymptotes are schematically shown in Fig.~\ref{figschem}$(b)$. The black line indicates the most accurate solution that utilizes non-local elasticity relation. It has the classical tip asymptotic behavior stemming from the problem of a semi-infinite plane strain hydraulic fracture (dashed red line). The dashed grey line indicates solution that employs the local elasticity together with the pressure boundary condition at the tip to capture the effect of fracture toughness. This leads to the discontinuous behavior of fracture width at the tip. Even though it is not physical, this boundary condition actually signifies the behavior at some distance from the tip $O(H)$. Indeed, as shown in~\cite{Dont2016a}, the solutions with local and non-local elasticity approach each other further away from the fracture tip, while the differences are concentrated in the $O(H)$ region near the tip. The global fracture characteristics, such as length and width at the wellbore are captured accurately, as soon as $H\ll l$. Further, one may also consider the tip asymptotics for PKN fracture, which is schematically shown by the dashed blue line. It is valid at the distances that are commensurate with the fracture length $l$, which is in contrast to the classical asymptote (dashed red line) that is accurate at the distances on the order of the fracture height $H$ for the geometry under consideration. To simplify the analysis, the model with local elasticity is considered in this study.

By using classical assumptions of PKN model~\cite{Perk1961,Nord1972}, each vertical fracture cross-section is assumed to be elliptical, and the pressure is determined based on the local plane strain elasticity assumption. This implies that
\begin{equation}\label{PKNrel}
w(x,y)~=~\dfrac{4}{\pi} \bar w(x) \sqrt{1-\Bigl(\dfrac{2y}{H}\Bigr)^2},\qquad p(x)~=~\dfrac{2 E' \bar w(x)}{\pi H},\qquad \bar w(x)=\dfrac{1}{H}\int_{-H/2}^{H/2} w(x,y)\,dy,
\end{equation}
where $w(x,y)$ denotes fracture opening, $\bar w(x)$ is the effective width, $p(x)$ is the fluid pressure, and $E'=E/(1\!-\!\nu^2)$ is the plane strain Young's modulus. Given the assumption $H\ll l$, the vertically-averaged lubrication equation for the case of Newtonian fracturing fluid becomes
\begin{equation}\label{lubrication}
\dfrac{\partial \bar w}{\partial t}+\dfrac{\partial \bar{q}_x }{\partial x} + \dfrac{C'}{\sqrt{t\!-\!t_0(x)}}~=~\dfrac{Q_0}{H} \delta(x),\qquad \bar{q}_x~=~-\frac{1}{12H\mu} \dfrac{\partial p}{\partial x} \int_{-H/2}^{H/2}w^3\, dy~=~-\frac{\bar w^3}{\pi^2\mu} \dfrac{\partial p}{\partial x},
\end{equation}
where $\mu$ is fluid viscosity, $\bar q_x$ is the average horizontal flux, $C'=2C_l$ is the scaled Carter's leak-off coefficient~\cite{Cart1957}, and $t_0(x)$ denotes the time instant at which the fracture front was located at the point $x$. Equations~(\ref{PKNrel}) and~(\ref{lubrication}) can be combined to yield
\begin{equation}\label{lubrication2}
\dfrac{\partial \bar w}{\partial t} - \dfrac{E'}{2\pi^3 \mu H}\dfrac{\partial ^2\bar w^4} {\partial x^2}+\dfrac{C'}{\sqrt{t\!-\!t_0(x)}}~=~\dfrac{Q_0}{H} \delta(x),
\end{equation}
which is the single governing equation for the PKN model.

To model lateral propagation condition that accounts for fracture toughness, the model from~\cite{Sarva2015} is employed, namely
\begin{equation}\label{PKNpbc}
p(l)~=~\dfrac{2 K_{Ic}}{\sqrt{\pi H}},
\end{equation}
where $K_{Ic}$ is fracture toughness and $l$ is the fracture half-length. As shown in~\cite{Dont2016a} this propagation condition is able to adequately capture the effect of toughness on lateral fracture growth, even though the near tip behavior is less accurate. Equation~(\ref{PKNpbc}) can be rewritten using~(\ref{PKNrel}) as
\begin{equation}\label{PKNwbc}
\bar w(l)~=~\dfrac{\sqrt{\pi H}\, K_{Ic}}{E'}.
\end{equation}
This form of the propagation condition is used as a boundary condition for~(\ref{lubrication2}), which makes the whole problem formulated only in terms of $\bar w(x)$ and $l(t)$. Pressure, on the other hand, can be obtained from~(\ref{PKNrel}) if needed.

The evolution of length is governed by the volume balance at the tip, which requires that
\begin{equation}\label{lengtheq}
    \dfrac{dl}{dt} = \dfrac{\bar q_x(l)}{\bar w(l)} = -\dfrac{2 E'}{3\pi^3 \mu H}\dfrac{\partial \bar w^3}{\partial x}\biggr|_{x=l},
\end{equation}
which relates the spatial derivative of the width to the rate of fracture growth.

For completeness, global volume balance can be obtained by integrating~(\ref{lubrication2}) with respect to time and space as
\begin{equation}\label{volbal}
\int_0^l \Bigl[\bar w(x) + 2C'\sqrt{t\!-\!t_0(x)}\Bigr] \,dx~=~\dfrac{Q_0t}{2H}.
\end{equation}
In the above expression the first term signifies fracture volume, the second term is the total leaked volume, while the right hand side is the injected volume.

\section{Tip region}\label{sectiontip}

The problem of a tip region is considered first, which corresponds to the semi-infinite PKN fracture propagating steadily with the velocity $V$. In order to derive the relevant governing equations for the tip region from~(\ref{lubrication2}), a standard approach is taken, whereby a new moving coordinate is introduced $\hat x = Vt\!-\!x$ (see e.g.~\cite{Gara2011,Peir2008}). In this case, the governing differential equation for the tip region becomes
\begin{equation}\label{PKNtip}
 \dfrac{E'}{2\pi^3 \mu H}\dfrac{d \bar w^4} {d \hat x}= V \bar w +2C'\sqrt{V\hat x},\qquad \bar w(0)~=~\dfrac{\sqrt{\pi H}\, K_{Ic}}{E'}.
\end{equation}
Here the second equation is simply the boundary condition~(\ref{PKNwbc}). 

Unfortunately, it is not possible to solve the differential equation~(\ref{PKNtip}) analytically. At the same time, one can obtain limiting solutions. In the limit of large toughness, the width is simply equal to the boundary condition in~(\ref{PKNtip}). To obtain the solution for zero toughness and zero leak-off (viscosity-dominated), both toughness and leak-off need to be set to zero in~(\ref{PKNtip}). This results in a solvable differential equation. In the limit of high leak-off and no toughness, the term ``$V\bar w$'' and toughness should be removed in~(\ref{PKNtip}), which again leads to the solvable differential equation. The summary of all three limiting or vertex solutions is
\begin{equation}\label{tipvertex}
    \bar w_k = \dfrac{\sqrt{\pi H}\, K_{Ic}}{E'},\qquad \bar w_m = \Bigl( \dfrac{3\pi^3\mu HV}{2E'}\Bigr)^{1/3}\,\hat x^{1/3},\qquad \bar w_{\tilde m} = \Bigl(\dfrac{8\pi^3\mu H C' V^{1/2}}{3E'}\Bigr)^{1/4}\, \hat x^{3/8},
\end{equation}
where the first is the toughness solution, the second is the viscous solution, and the third is the leak-off solution. These results agree with the previous analysis in~\cite{Koval2010,Sarva2015}.

It is also possible to compute the $km$ and $k\tilde m$ solutions. Indeed, the differential equation~(\ref{PKNtip}) can be solved for these cases to obtain
\begin{equation}\label{tipedges}
    \bar w_{km}  = \bigl(\bar w_k^3 + \bar w_m^3 \bigr)^{1/3},\qquad \bar w_{k\tilde m} = \bigl(\bar w_k^4+\bar w_{\tilde m}^4\bigr)^{1/4}.
\end{equation}
Motivated by such a solution form, the following approximate solution is constructed
\begin{equation}\label{tipglobalapproxsimple}
    \bar w_{m\tilde mk}  = \bigl(\bar w_k^p + \bar w_m^p +\bar w_{\tilde m}^p\bigr)^{1/p},\qquad p=3.4,
\end{equation}
where the power $p$ is tuned to minimize the maximum error with the numerical solution. This solution captures the limiting vertex cases precisely, while approximates the behavior in the transition regions. It is also possible to construct a more accurate solution:
\begin{equation}\label{tipglobalapprox}
    \bar w_{m\tilde mk}  = \Bigl[w_{km}\bigl(\bar w_{km}^4 + \bar w_{\tilde m}^4\bigr)^{1/4}+w_{k\tilde m}\bigl(\bar w_{k\tilde m}^3 + \bar w_{m}^3\bigr)^{1/3}\Bigr] \bigl[w_{km}+w_{k\tilde m} \bigr]^{-1},
\end{equation}
which in addition also respects the edge solutions outlined above.

In order to check accuracy of the suggested approximations and for simpler analysis, it is convenient to reformulate the problem~(\ref{PKNtip}) in the dimensionless form by introducing the following parameters:
\begin{equation}
    \Omega = \dfrac{\bar w}{\bar w_k} = \dfrac{ E' \bar w}{(\pi H)^{1/2} K_{Ic}},\qquad \hat\xi = \dfrac{ \pi^{3/2}\mu V E'^2 \hat x}{2 K_{Ic}^3 H^{1/2}},\qquad \chi = \Bigl(\dfrac{8C'^2K_{Ic}}{\pi^{5/2}\mu H^{1/2} V^2}\Bigr)^{1/2}.
\end{equation}
In this case, equation~(\ref{PKNtip}) reduces to
\begin{equation}\label{PKNtipdim}
 \dfrac{d \Omega} {d \hat\xi}= \dfrac{1}{\Omega^{2}} +\dfrac{\chi \hat \xi^{1/2}}{\Omega^{3}},\qquad \Omega(0)~=~1.
\end{equation}
At the same time, the vertex solutions become
\begin{equation}\label{tipvertexdim}
    \Omega_{k}  = 1,\qquad \Omega_m = (3\hat\xi)^{1/3},\qquad \Omega_{\tilde m} = \Bigl(\dfrac{8\chi}{3}\Bigr)^{1/4}\hat\xi^{3/8},
\end{equation}
which also allows to recast the two suggested approximations~(\ref{tipglobalapproxsimple}) and~(\ref{tipglobalapprox}).

\begin{figure}[h]
\centering\includegraphics[width=0.95\linewidth]{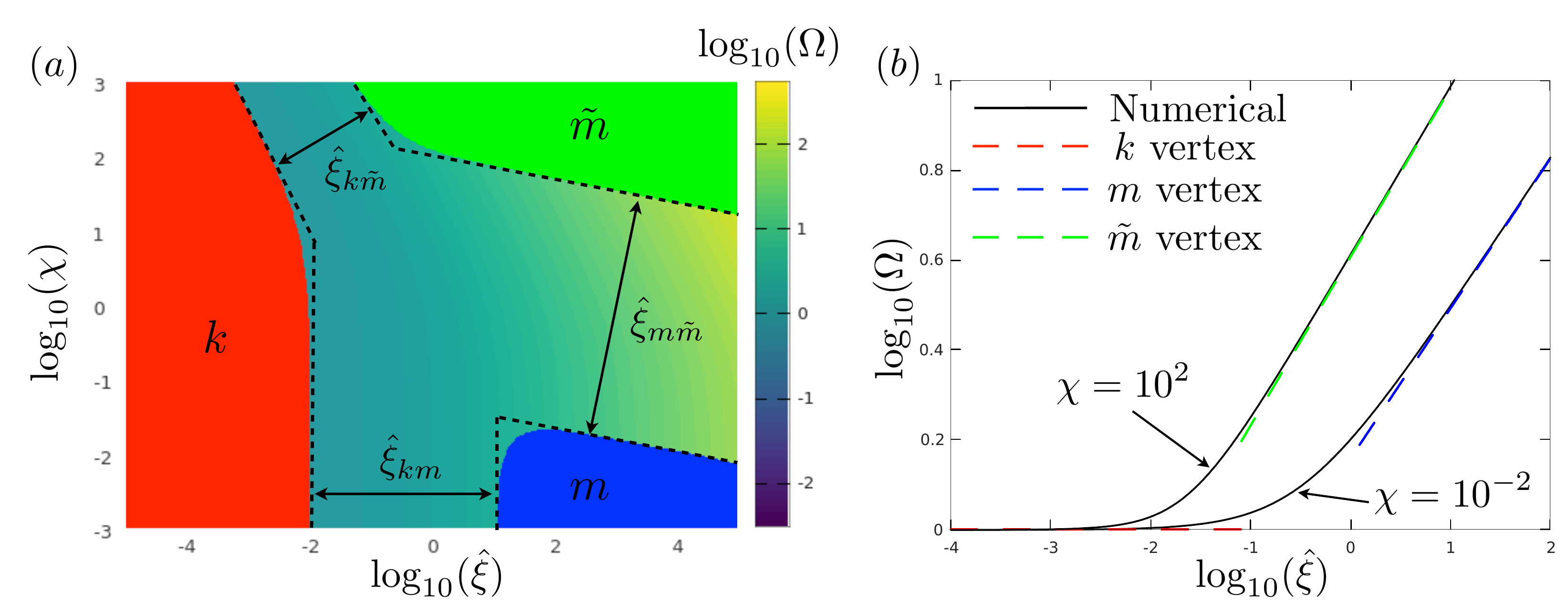}
\caption{Panel $(a)$: Parametric space for the problem of tip region for PKN fracture. The colored regions indicate zones of applicability of the limiting vertex solutions~(\ref{tipvertexdim}) and they are also outlined by the dashed black lines defined by~(\ref{tiptransitions}). Panel $(b)$: Numerical solution for the dimensionless opening versus dimensionless coordinate for two values of leak-off $\chi=10^{-2}$ and $\chi=10^2$ (solid black lines). The vertex solutions~(\ref{tipvertexdim}) are shown by the dashed colored lines.}
\label{figPKNtipmap}
\end{figure}

Fig.~\ref{figPKNtipmap}$(a)$ plots the numerical solution of~(\ref{PKNtipdim}) (computed using Runge-Kutta 4th order explicit integration of~(\ref{PKNtipdim})) in the parametric space $(\hat\xi,\chi)$. Regions of applicability of the vertex solutions~(\ref{tipvertexdim}) are highlighted by the colored regions. These regions are defined as the zones, in which the relative difference between the global solution and the corresponding limiting solution does not exceed 1\%. In particular, the red region corresponds to the applicability of the toughness or ``$k$'' solution, the blue region represents the viscosity or ``$m$'' solution, while the green zone shows the zone of applicability of the leak-off or ``$\tilde m$'' solution. There are three transition regions, namely from $k$ to $m$, from $k$ to $\tilde m$, and from $m$ to $\tilde m$. These are quantified by the following parameters:
\begin{eqnarray}\label{tiptransitions}
\hat\xi_{km}&=&\hat\xi,\qquad \hat\xi_{km,1} \approx 0.010,\qquad \hat\xi_{km,2} \approx 11,\notag\\
\hat\xi_{k\tilde m}&=&\hat\xi\chi^{2/3},\qquad \hat\xi_{k\tilde m,1} \approx 0.062,\qquad \hat\xi_{k\tilde m,2} \approx 4.4,\\
\hat\xi_{m\tilde m}&=&\hat\xi\chi^{6},\qquad \hat\xi_{m\tilde m,1} \approx 1.85\times 10^{-8},\qquad \hat\xi_{m\tilde m,2} \approx 2.1\times 10^{12}.\notag
\end{eqnarray}
These parameters, alongside with their respective minimum and maximum values, quantitatively determine boundaries of applicability of the limiting vertex solutions and are plotted by the dashed black lines in Fig.~\ref{figPKNtipmap}$(a)$. The simplest way to determine these transitional parameters is to equate the corresponding vertex solutions scale-wise, i.e. $\Omega_k\sim\Omega_{\tilde m}$ or $1\sim \chi^{1/4}\hat\xi^{3/8}$, which implies that $\xi_{k\tilde m} = \hat\xi \chi^{2/3}$.

Fig.~\ref{figPKNtipmap}$(b)$ shows the linear plots of the numerical solution computed for $\chi=10^{-2}$ and $\chi=10^2$ versus $\hat\xi$ (black lines). The dashed colored lines indicate the vertex solutions~(\ref{tipvertexdim}). Gradual transitions from the toughness to the leak-off or viscosity solutions are observed. 

One peculiar observation is that the solution starts from the $k$ vertex, then for some values of $\chi$ passes through the intermediate asymptote $m$, while $\tilde m$ is the far-field solution. This is in contrast to the classical solution for a semi-infinite plane strain hydraulic fracture~\cite{Gara2011}, for which the far-field is always $m$ solution, while $\tilde m$ is intermediate. Also, given the relatively gentle slope of the $m\tilde m$ transition in the parametric space Fig.~\ref{figPKNtipmap}$(a)$, the transition from $m$ to $\tilde m$ can be hard to achieve in practice.

\begin{figure}[h]
\centering\includegraphics[width=0.95\linewidth]{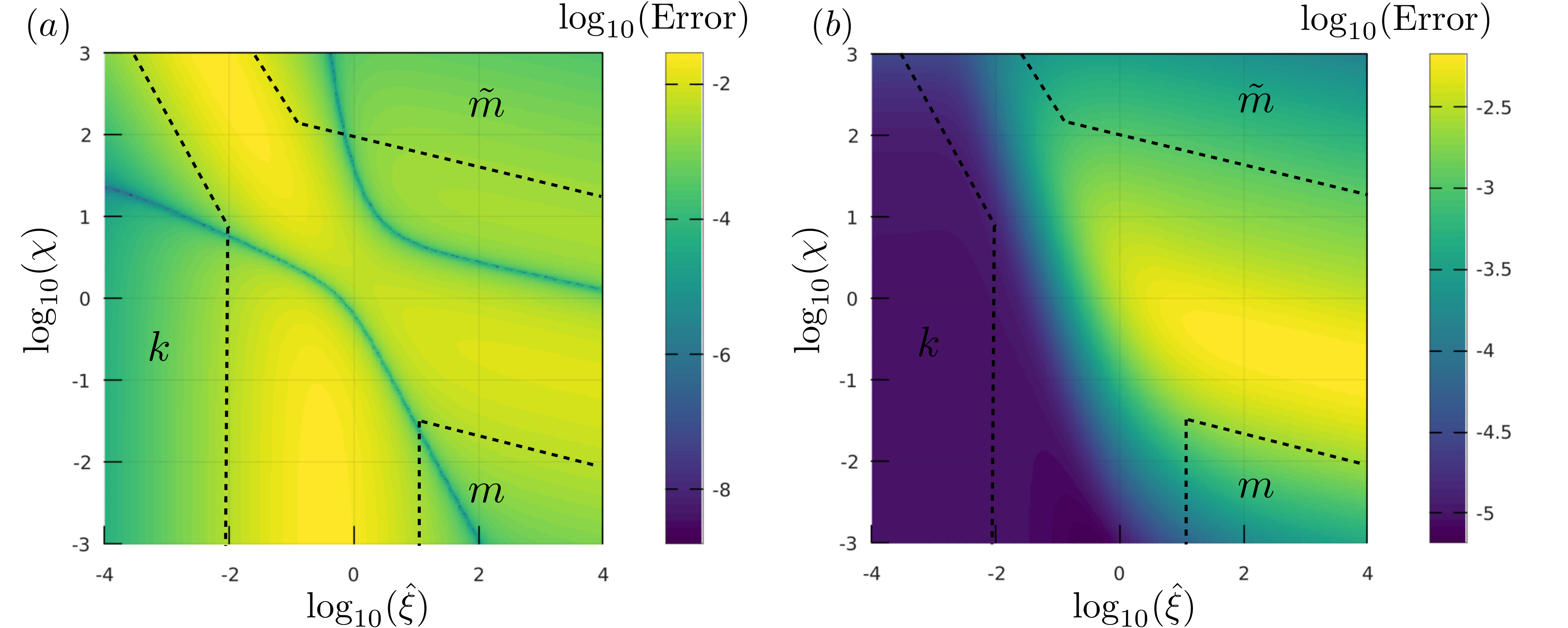}
\caption{Relative error between the numerical solution and approximations~(\ref{tipglobalapproxsimple}) (panel $(a)$) and~(\ref{tipglobalapprox}) (panel $(b)$).}
\label{figPKNtiperr}
\end{figure}
In order to quantify accuracy of the approximations~(\ref{tipglobalapproxsimple}) and~(\ref{tipglobalapprox}), Fig.~\ref{figPKNtiperr} plots the relative error between the approximate and numerical solutions. Panel $(a)$ gives the result for the simple solution~(\ref{tipglobalapproxsimple}), while panel $(b)$ corresponds to the more comprehensive approximation~(\ref{tipglobalapprox}). The maximum error within the entire parametric space for the first case is 3\%, while it is just below 0.7\% for the second, more accurate, case. The distribution of the error is different as well. For the simpler case~(\ref{tipglobalapproxsimple}), the error is distributed roughly uniformly within the transition regions, and decays towards the vertex limits. For the second case~(\ref{tipglobalapprox}), the $k$ vertex and the transitions from this vertex to $m$ and $\tilde m$ are noticeably more accurate, while the error occurs mostly within the $m\tilde m$ transition. This is not surprising, since the exact solutions for $km$ and $k\tilde m$ transitions are used to construct the approximation~(\ref{tipglobalapprox}).

\section{Vertex solutions}\label{sectionvertex}

Similarly to the problems of a plane strain and radial fracture, there are four limiting solutions for the PKN geometry. These are determined by the competition between toughness and viscosity, as well as storage of fluid within fracture and porous formation. In particular, let's define storage viscosity or $M$ limit as the case corresponding to domination of viscosity and no leak-off, the storage toughness or $K$ limit as the case of toughness domination and no leak-off, the leak-off viscosity or $\tilde M$ limit as the case of viscosity domination and high leak-off, and finally leak-off toughness or $\tilde K$ limit as the case of toughness domination and high leak-off, see e.g. the review paper~\cite{Detou2016}.

\paragraph{Storage viscosity.} For the storage viscosity case, the governing equation~(\ref{lubrication2}), the boundary condition~(\ref{PKNwbc}), and the volume balance~(\ref{volbal}) reduce to
\begin{equation}\label{govM}
\dfrac{\partial \bar w}{\partial t} - \dfrac{E'}{2\pi^3 \mu H}\dfrac{\partial ^2\bar w^4} {\partial x^2}~=~\dfrac{Q_0}{H}\delta(x),\qquad
\int_0^l \bar w(x) \,dx~=~\dfrac{Q_0t}{2H},\qquad \bar w(l)=0.
\end{equation}
By explicitly incorporating the viscosity dominated near-tip behavior~(\ref{tipvertex}) into the global solution and looking for the solution in the form
\begin{equation}\label{wM}
    \bar w=\Bigl(\dfrac{3\pi^3\mu H} {2E'}\Bigr)^{1/3} (\dot l)^{1/3} l^{1/3} (1\!-\!\xi)^{1/3} f(\xi),\qquad \xi=\dfrac{x}{l(t)},
\end{equation} 
the volume balance in~(\ref{govM}) reduces to
\begin{equation}\label{govM2}
\Bigl(
\dfrac{3\pi^3\mu H}{2E'}\Bigr)^{1/3} (\dot l)^{1/3} l^{4/3}\int_0^1 (1\!-\!\xi)^{1/3}f(\xi) \,d\xi~=~\dfrac{Q_0t}{2H}.
\end{equation}
By balancing the powers of $t$ and other dimensional parameters in the above equation, solution for length can be written as
\begin{equation}\label{Mlength}
l=c\,\Bigl(\dfrac{5E' Q_0^3 }{48\pi^3 \mu H^{4}}\Bigr)^{1/5} t^{4/5},
\end{equation}
where $c$ is a numeric constant and the volume balance equation becomes
\begin{equation}\label{govM3}
c^{5/3}\int_0^1 (1\!-\!\xi)^{1/3}f(\xi) \,d\xi~=~1.
\end{equation}
With the above definition for width~(\ref{wM}) and length~(\ref{Mlength}), the lubrication equation in~(\ref{govM}) transforms to
\begin{equation}\label{govM4}
\dfrac{1}{4}(1\!-\!\xi)^{1/3} f(\xi) -  \xi \dfrac{d (1\!-\!\xi)^{1/3} f(\xi)}{d\xi}  - 
\dfrac{3}{4}
\dfrac{d^2 (1\!-\!\xi)^{4/3}(f(\xi))^4} {d \xi^2}~=~0,
\end{equation}
where the boundary conditions become
\begin{equation}\label{bcdim}
   - c^{5/3}\,\dfrac{3}{5}\dfrac{d (1\!-\!\xi)^{4/3} f^4} {d\xi}\Bigr|_{\xi=0} = 1,\qquad f\bigr|_{\xi=1}=1.
\end{equation}
Instead of solving the differential equation~(\ref{govM4}), the solution is sought in the form $f=1+a(1\!-\!\xi)$, i.e. expanding it near the tip. By substituting this result into~(\ref{govM4}) and balancing the powers of $(1\!-\!\xi)$ assuming that the latter is a small parameter, the result is $a=-1/96$. This demonstrates that the linear correction to the tip solution (i.e. $f=1$) is on the order of 1\%, even near the source. Therefore, it seems sufficient to consider only the linear term and write the approximate solution for $f$ as
\begin{equation}\label{fsol}
    f(\xi)\approx1-\dfrac{1\!-\!\xi}{96}.
\end{equation}
By substituting~(\ref{fsol}) into~(\ref{govM3}), the numeric coefficient $c$ becomes
\begin{equation}\label{csol}
    c=\Bigl(\dfrac{224}{167}\Bigr)^{3/5}\approx 1.193.
\end{equation}
To summarize, the approximate $M$ vertex solution is computed using (\ref{wM}),  (\ref{Mlength}), (\ref{fsol}), and~(\ref{csol}) as
\begin{eqnarray}\label{solMPKN}
  \bar w_{M}&=& 1.76 \,\Bigl(\dfrac{ \mu Q_0^2}{E' H}\Bigr)^{1/5} t^{1/5} (1\!-\!\xi)^{1/3}\Bigl(1-\dfrac{1\!-\!\xi}{96}\Bigr),\notag \\
  p_{M}&=&1.12\,\Bigl(\dfrac{ \mu E'^4Q_0^2}{ H^6}\Bigr)^{1/5} t^{1/5}(1\!-\!\xi)^{1/3}\Bigl(1-\dfrac{1\!-\!\xi}{96}\Bigr),\\
   l_{M}&=&0.38\,\Bigl(\dfrac{E' Q_0^3 }{ \mu H^{4}}\Bigr)^{1/5} t^{4/5}.\notag
\end{eqnarray}
Here the local elasticity equation~(\ref{PKNrel}) is used to compute pressure. Note that this approximate solution is based primarily on the behavior near the tip and the condition at the inlet~(\ref{bcdim}) is not satisfied. Therefore, there can be some discrepancies with the ``true'' solution near the inlet. These expressions match that outlined in~\cite{Econo2000} to within a few percent error. This is an acceptable difference since approximations are also used to derive the expressions presented in~\cite{Econo2000}.




\paragraph{Leak-off viscosity.} 

For the leak-off viscosity case, both toughness and fracture volume are negligibly small. As a result, the governing equation~(\ref{lubrication2}), the boundary condition~(\ref{PKNwbc}), and the global volume balance~(\ref{volbal}) reduce to
\begin{equation}\label{govMt}
- \dfrac{E'}{2\pi^3 \mu H}\dfrac{\partial ^2\bar w^4} {\partial x^2}+\dfrac{C'}{\sqrt{t\!-\!t_0(x)}}~=~\dfrac{Q_0}{H}\delta(x),\qquad
2C'\int_0^l \sqrt{t\!-\!t_0(x)}\,dx~=~\dfrac{Q_0t}{2H},\qquad \bar w(l)=0.
\end{equation}
By taking the time dependence of length as $l\propto t^\alpha$, the exposure time needed for leak-off calculations becomes $t_0(x)=t(x/l)^{1/\alpha}$. In this case, the global volume balance reduces to
\begin{equation}\label{volbalMt2}
2C't^{1/2} l \int_0^1 \sqrt{1\!-\!\xi^{1/\alpha}} \,d\xi~=~\dfrac{Q_0t}{2H}.
\end{equation}
By balancing the powers of $t$ in the above expression, one has $\alpha = 1/2$ and the length is
\begin{equation}\label{lMt}
     l=\dfrac{Q_0 t^{1/2} }{\pi C' H}.
\end{equation}
Knowing the expression for $t_0(x)$, the lubrication equation in~(\ref{govMt}) reduces to
\begin{equation}\label{govMt2}
- \dfrac{E'}{2\pi^3 \mu H}\dfrac{\partial ^2\bar w^4} {\partial x^2}+\dfrac{C'}{\sqrt{t} \sqrt{1-(x/l)^2}}~=~\dfrac{Q_0}{H}\delta(x),\qquad \bar w(l)=0,
\end{equation}
which can be solved to obtain
\begin{equation}\label{govMtsol}
\bar w = \Bigl(\dfrac{2\pi \mu Q_0^2}{E' C' H} \Bigr)^{1/4} t^{1/8} g(\xi),\qquad g(\xi) = \Bigl[\xi\Bigl(\sin^{-1}\bigl(\xi\bigr)-\dfrac{\pi}{2}\Bigr)+\sqrt{1-\xi^2}\Bigr]^{1/4},\qquad \xi=\dfrac{x}{l}.
\end{equation}
Finally, the solution for $\tilde M$ vertex can be summarized as
\begin{eqnarray}\label{solMtPKN}
  \bar w_{\tilde M}&=& \Bigl(\dfrac{2\pi \mu Q_0^2}{E' C' H} \Bigr)^{1/4} t^{1/8} g(\xi),\notag \\
  p_{\tilde M}&=&\Bigl(\dfrac{32 \mu E'^3 Q_0^2}{\pi^3 C' H^5} \Bigr)^{1/4} t^{1/8} g(\xi),\\
   l_{\tilde M}&=&\dfrac{Q_0 t^{1/2} }{\pi C' H}.\notag
\end{eqnarray}
where the function $g(\xi)$ is defined in~(\ref{govMtsol}). This solution coincides precisely with the one in~\cite{Econo2000}.

Note that the function $g(\xi)$ defined in~(\ref{govMtsol}) has the following asymptotic behavior
\begin{equation}
g(\xi)|_{\xi\rightarrow 1} = \Bigl(\dfrac{8}{9}\Bigr)^{1/8} (1-\xi)^{3/8},\qquad g(0) = 1.
\end{equation}
Given that $(8/9)^{1/8}\approx 0.985$, i.e. close to 1, the function $g$ can be approximated via its asymptotic behavior as
\begin{equation}\label{gapprox}
g(\xi)\approx\Bigl(\dfrac{8}{9}\Bigr)^{1/8} (1-\xi)^{3/8}\Bigl[1-\Bigl(1\!-\!\Bigl(\dfrac{9}{8}\Bigr)^{1/8} \Bigr)(1\!-\!\xi)\Bigr].
\end{equation}
This form of the solution for fracture width is very similar to that for the viscosity storage solution~(\ref{solMPKN}). What is also common is that the multiplier in front of the ``correction term'' $1\!-\!\xi$ is relatively small, on the order of 1\%.

\paragraph{Storage toughness.} In the limit of no viscosity and no leak-off, all the injected volume stays inside the fracture and there is no pressure gradient along the fracture. As a consequence of constant pressure and local elasticity, the width is constant throughout the fracture in this limit. The combination of the boundary condition~(\ref{PKNwbc}), and the volume balance~(\ref{volbal}) leads to the solution in the form
\begin{eqnarray}\label{solKPKN}
  \bar w_{K}&=& \dfrac{K_{Ic}\sqrt{\pi H}}{E'},\notag \\
  p_{K}&=& \dfrac{2K_{Ic}}{\sqrt{\pi H}},\\
   l_{K}&=&\dfrac{E' Q_0 t }{ \sqrt{4\pi} K_{Ic} H^{3/2}}.\notag
\end{eqnarray}

\paragraph{Leak-off toughness.} In the limit of toughness domination, the width and pressure are constant along the fracture and are determined solely by the boundary condition at the tip~(\ref{PKNwbc}), just like for the previous case. The difference comes from the volume balance, in which the injected fluid is now balanced by leak-off, namely
\begin{equation}\label{volbalKt}
\int_0^l 2C'\sqrt{t\!-\!t_0(x)} \,dx~=~\dfrac{Q_0t}{2H}.
\end{equation}
Similarly to the case of leak-off viscosity, by taking the time dependence of length as $l\propto t^\alpha$, one has $t_0(x)=t(x/l)^{1/\alpha}$, in which case the above integral reduces to
\begin{equation}\label{volbalKt2}
2C't^{1/2} l \int_0^1 \sqrt{1\!-\!\xi^{1/\alpha}} \,d\xi~=~\dfrac{Q_0t}{2H}.
\end{equation}
By balancing the powers of $t$, it is now clear that $\alpha = 1/2$. As a result, the above equation allows to find $l$, which, combined with the expressions for width and pressure stemming from the boundary condition, gives the following result
\begin{eqnarray}\label{solKtPKN}
  \bar w_{\tilde K}&=& \dfrac{K_{Ic}\sqrt{\pi H}}{E'},\notag \\
  p_{\tilde K}&=& \dfrac{2K_{Ic}}{\sqrt{\pi H}},\\
   l_{\tilde K}&=&\dfrac{Q_0 t^{1/2} }{\pi C' H}.\notag
\end{eqnarray}
It is interesting to observe that the difference of this result with the previous toughness storage solution is just in the fracture length, while the width and the pressure are exactly the same. The length, on the other hand is the same as for the viscosity leak-off limit.

\section{Full solution}\label{sectionfull}

This section aims to construct and analyze the full solution for the problem of PKN fracture, which includes the effects of toughness, viscosity, and leak-off. Two types of solutions are constructed: semi-analytical rapid approximate solution and fully numerical solution. The former allows to compute solution quickly and to investigate the whole parametric space, while the latter allows to check accuracy of the approximation.

As can be seen from the limiting vertex solutions~(\ref{solMPKN}), (\ref{solMtPKN}), (\ref{solKPKN}), and~(\ref{solKtPKN}), spatial variation of the fracture width is closely approximated by the corresponding tip asymptotes~(\ref{tipvertex}). There is some discrepancy for the storage viscosity and leak-off viscosity solutions, but it is on the order of 1\% and is therefore ignored for simplicity. Motivated by this observation, the width of the global solution is taken in the form:
\begin{equation}\label{approxbarw}
    \bar w = \bar w_{m\tilde m k}(\hat x=l(1\!-\!\xi),V=\dot l),\qquad \xi=\dfrac{x}{l},
\end{equation}
where $\bar w_{m\tilde m k}$ is the approximate solution for the semi-infinite fracture~(\ref{tipglobalapprox}), which also depends on $E'$, $K_{Ic}$, $\mu$, $C'$, and $H$. Note that the approach for constructing the global solution from the tip asymptote was applied in~\cite{Dont2016f} for a radial hydraulic fracture and~\cite{Dont2017c} for a plane strain hydraulic fracture. One notable difference for PKN fracture is the fact that the tip asymptote is discontinuous at the tip and therefore it is not very accurate to approximate the spatial behavior by $(1\!-\!\xi)^\delta$ (with slowly varying function $\delta$), as was done for the prior cases. In order to construct the global solution using~(\ref{approxbarw}), it is also necessary to consider global volume balance~(\ref{volbal}). By taking $l\propto t^\alpha$, where $\alpha$ is a slowly varying function of time, the leak-off exposure time becomes $t_0(x)=t(x/l)^{1/\alpha}$ and the volume balance~(\ref{volbal}) reduces to
\begin{equation}\label{volbal2}
l \int_0^1 \bar w_{m\tilde mk}(l(1\!-\!\xi),\alpha l/t)\,d\xi + 2C't^{1/2} l\int_0^1\sqrt{1\!-\!\xi^{1/\alpha}} \,d\xi~=~\dfrac{Q_0t}{2H}.
\end{equation}
The leak-off related integral in the above equation can be computed analytically, which results in the following system of equations
\begin{equation}\label{apprsolsys}
l \int_0^1 \bar w_{m\tilde mk}(l(1\!-\!\xi),\alpha l/t)\,d\xi+ \sqrt{\pi}C't^{1/2} l\dfrac{ \Gamma(\alpha+1)}{\Gamma(\alpha+\tfrac{3}{2})}~=~\dfrac{Q_0t}{2H},\qquad \alpha = \dfrac{d\log(l)}{d\log(t)},
\end{equation}
where $\Gamma(\cdot)$ denotes Gamma function. The integral of the fracture width in the above expression is computed numerically. Since $\alpha$ varies slowly, the first equation can solved iteratively for $l$, e.g. using Newton's method. Once the length $l$ computed for two time instants, the value of $\alpha$ is updated. Such iterations are continued until convergence is reached. Once these quantities are computed, then the width solution is evaluated using~(\ref{approxbarw}), while pressure can always be computed with the help of~(\ref{PKNrel}). 

To compute numerical solution for the problem under consideration, similarly to the previous case, the moving coordinate $\xi=x/l(t)$ is introduced. To further simplify the problem, the following normalization is employed
\begin{equation}\label{normalization}
    \Omega = \dfrac{\bar w}{w_*},\qquad \lambda = \dfrac{l}{l_*},\qquad \tau = \dfrac{t}{t_*},
\end{equation}
where the width, length, and time scales are given by
\begin{equation}\label{scales}
    w_* = \dfrac{(\pi H)^{1/2}\, K_{Ic}}{E'},\qquad l_* = \dfrac{H^2 K_{Ic}^4}{2\pi E'^3\mu Q_0} ,\qquad t_* = \dfrac{H^{7/2} K_{Ic}^5}{2 \pi^{1/2} E'^4\mu Q_0^2}.
\end{equation}
With such a normalization, the governing equations~(\ref{lubrication2}), (\ref{PKNwbc}) and~(\ref{lengtheq}) reduce to
\begin{equation}\label{PKNdim}
\dfrac{\partial \Omega}{\partial \tau}-\dfrac{\xi\dot\lambda}{\lambda}\dfrac{\partial \Omega}{\partial \xi} - \dfrac{1}{\lambda^2}\dfrac{\partial ^2 \Omega^4} {\partial \xi^2}+\dfrac{\phi}{\sqrt{\tau\!-\!\tau_0(\xi)}}~=~ \delta(\xi),\qquad \Omega(1) = 1,\qquad \dfrac{d\lambda}{d\tau} = -\dfrac{4}{3\lambda}\dfrac{\partial \Omega^3}{\partial \xi}\biggr|_{\xi=1}.
\end{equation}
Here the dimensionless parameter that describes leak-off is computed as
\begin{equation}\label{phidef}
    \phi = \Bigl(\dfrac{H^{5} K_{Ic}^{6} C'^4}{4 \pi^3 E'^4\mu^2 Q_0^4} \Bigr)^{1/4}.
\end{equation}
Solution for the above dimensionless problem~(\ref{PKNdim}) comprises of finding $\Omega(\tau,\phi)$ and $\lambda(\tau)$. Once the dimensionless problem is solved, the physical quantities can be restored using~(\ref{normalization}) and~(\ref{scales}). This demonstrates that the parametric space for the problem is two-dimensional and consists of the dimensionless time $\tau$, and dimensionless leak-off $\phi$, which is somewhat similar to the case of radial hydraulic fracture~\cite{Dont2016f}. Numerical solution for the problem~(\ref{PKNdim}) is computed using central difference to discretize spatial derivatives and backward difference for approximating the time derivative to ensure stability of the numerical scheme.

\begin{figure}[h]
\centering\includegraphics[width=0.5\linewidth]{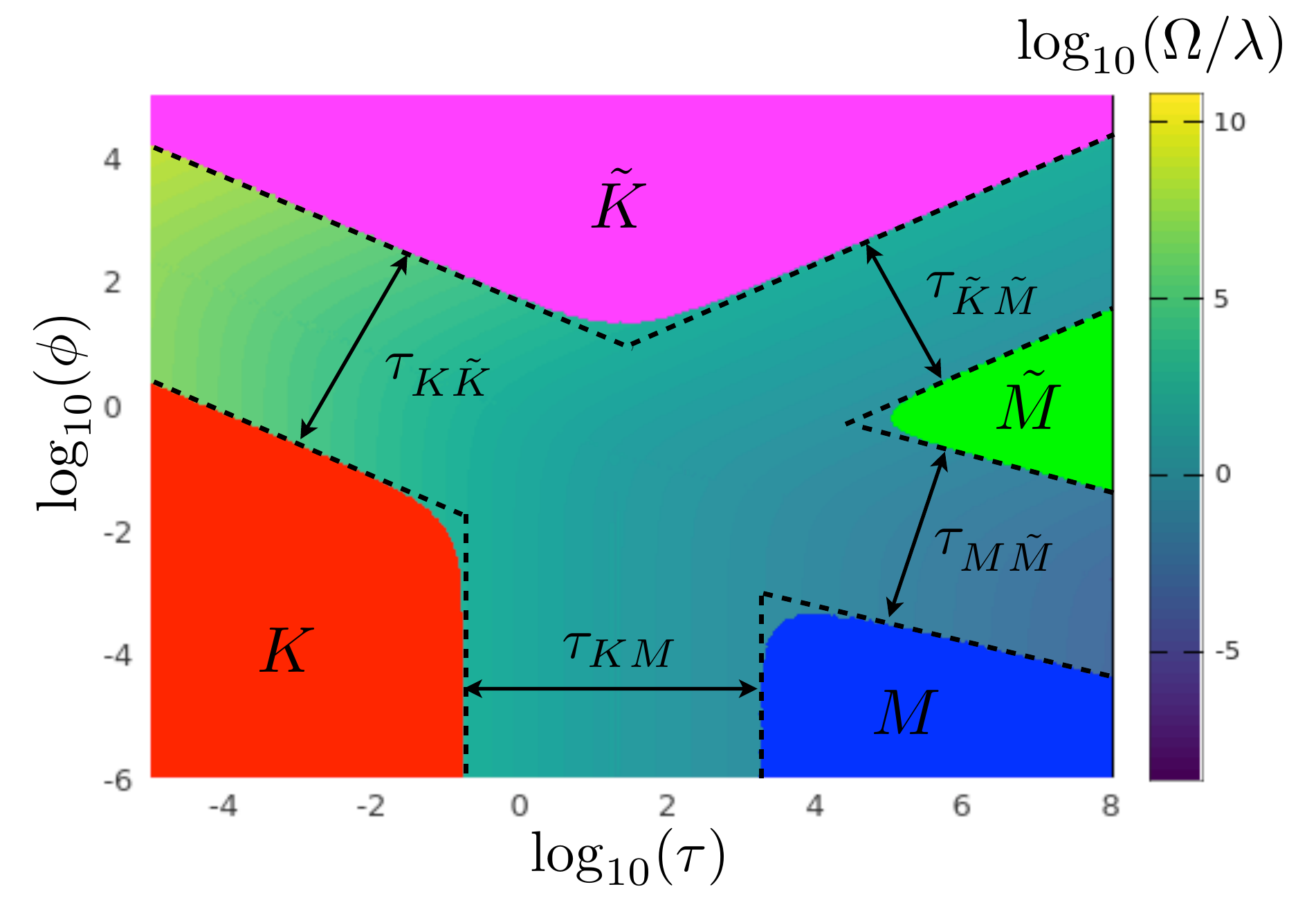}
\caption{Parametric space for PKN hydraulic fracture. Zones of applicability of the limiting vertex solutions are highlighted by the red ($K$ or storage toughness), blue ($M$ or storage viscosity), magenta ($\tilde K$ or leak-off toughness), and green ($\tilde M$ or leak-off viscosity).}
\label{figPKNparamspace}
\end{figure}

Fig.~\ref{figPKNparamspace} plots the parametric space for the problem that is computed using the fast approximate solution~(\ref{apprsolsys}). In particular, the solution is plotted in the $(\tau,\phi)$ space in terms of the ratio $\Omega/\lambda$ (see~(\ref{normalization}), (\ref{scales}), and~(\ref{phidef})). This is done to ensure that the plotted quantity is different for all the limiting cases. Zones of applicability of the limiting vertex solutions are indicated by the red, blue, magenta, and green areas. The latter are defined as the regions, in which the difference (of $\Omega/\lambda$) with the corresponding vertex solution is below 1\%. The red region corresponds to the storage toughness solution~(\ref{solKPKN}), the blue region represents the storage viscosity limit~(\ref{solMPKN}), the magenta region defines the leak-off toughness limit~(\ref{solKtPKN}), while the green region corresponds to the leak-off viscosity solution~(\ref{solMtPKN}). The dashed lines in Fig.~\ref{figPKNparamspace} outline zones of applicability of the vertex solutions. To obtain the relevant parameter that determines the transition, one needs to equate either lengths or widths (whichever are different) between two regimes. For instance, the transition between $K$ and $M$ is determined by the parameter calculated from $l_{M}\sim l_K$ or $\bar w_{M}\sim \bar w_K$. To quantify the parameter for $\tilde K K$ transition, one should use $l_{\tilde K}\sim l_K$. The widths cannot be used since they are identical. Similarly, for the $\tilde K \tilde M$ transition one has to use widths $\bar w_{\tilde K}\sim \bar w_{\tilde M}$ since the lengths are identical. Finally, for the $M\tilde M$ either the widths or the lengths can be used, i.e. $l_{M}\sim l_{\tilde M}$ or $\bar w_{M}\sim \bar w_{\tilde M}$. The values of the transition parameters are determined numerically. Summary of the results is shown below:
\begin{eqnarray}\label{PKNtransitions}
\tau_{MK} &=& \tau, \qquad\tau_{MK,1} = 0.11,\qquad\tau_{MK,2} = 2.3\times 10^3,\notag\\
\tau_{K\tilde K}&=& \tau \phi^{2},\qquad \tau_{K\tilde K,1} = 5.7\times 10^{-5},\qquad\tau_{K\tilde K,2} = 3.1\times 10^{3},\notag\\
\tau_{\tilde K \tilde M}&=&\tau \phi^{-2},\qquad \tau_{\tilde K\tilde M,1} = 0.18,\qquad\tau_{\tilde K\tilde M,2} = 6.5\times 10^{4},\notag\\
\tau_{M \tilde M} &=& \tau\phi^{10/3},\qquad \tau_{M\tilde M,1} = 2.0\times 10^{-7},\qquad\tau_{M\tilde M,2} = 2.9\times 10^{3}.
\end{eqnarray}
Note that the definitions of the applicability zones of the limiting solutions are somewhat arbitrary and the result can change if a different metric or different threshold is used. In addition, there can be some minor discrepancy since the approximate solution is used to construct the parametric map. At the same time, since the parametric space is plotted on a logarithmic scale, minor variations in the boundaries defined by~(\ref{PKNtransitions}) are not going to notably alter the result.
 
What is interesting about the obtained result is that the fracture evolves in time from the $K$ vertex and ultimately reaches the $\tilde M$ vertex. It can pass through $\tilde K$, $M$, or none of those during the intermediate times. This is somewhat opposite to the radial fracture geometry, for which the solution always originates from the $M$ vertex and then reaches the $\tilde K$ solution in the long term. It is also different from the plane strain fracture, for which the solution originates at the $MK$ edge and eventually reaches $\tilde M\tilde K$ edge.

\begin{figure}[h]
\centering\includegraphics[width=1.0\linewidth]{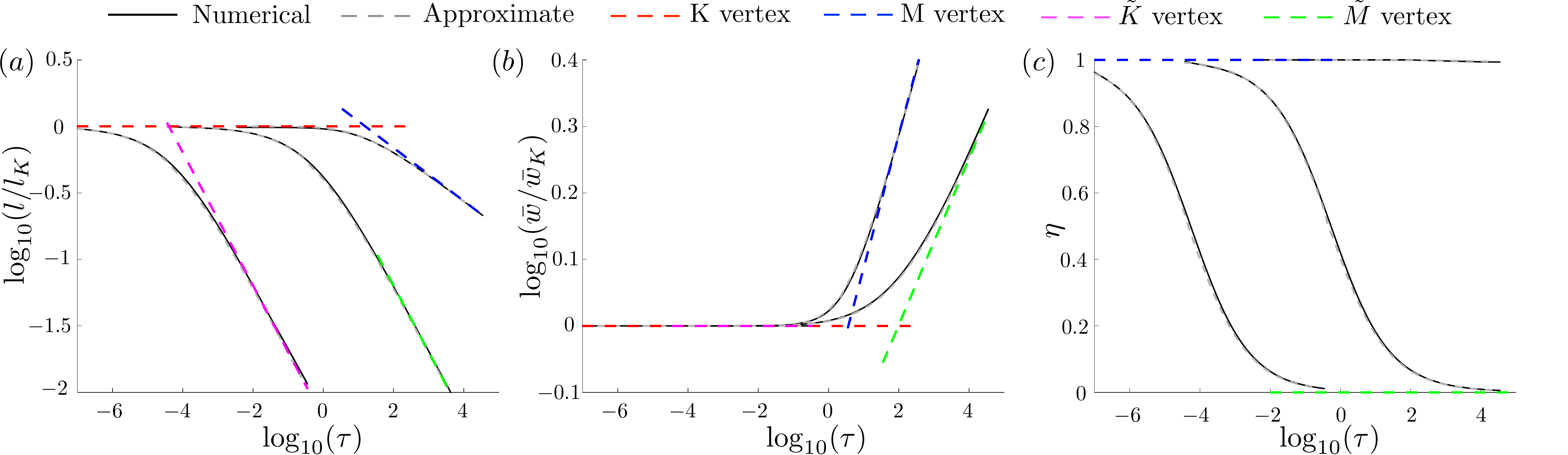}
\caption{Comparison between the numerical, approximate, and vertex solutions for three values of the dimensionless leak-off $\phi=\{10^{-4},1,10^2\}$. Panel $(a)$ shows the normalized fracture length versus time. Panel $(b)$ shows the normalized width at the wellbore versus time. Panel $(c)$ shows efficiency versus time.}
\label{figlweta}
\end{figure}
\begin{figure}[h]
\centering\includegraphics[width=0.75\linewidth]{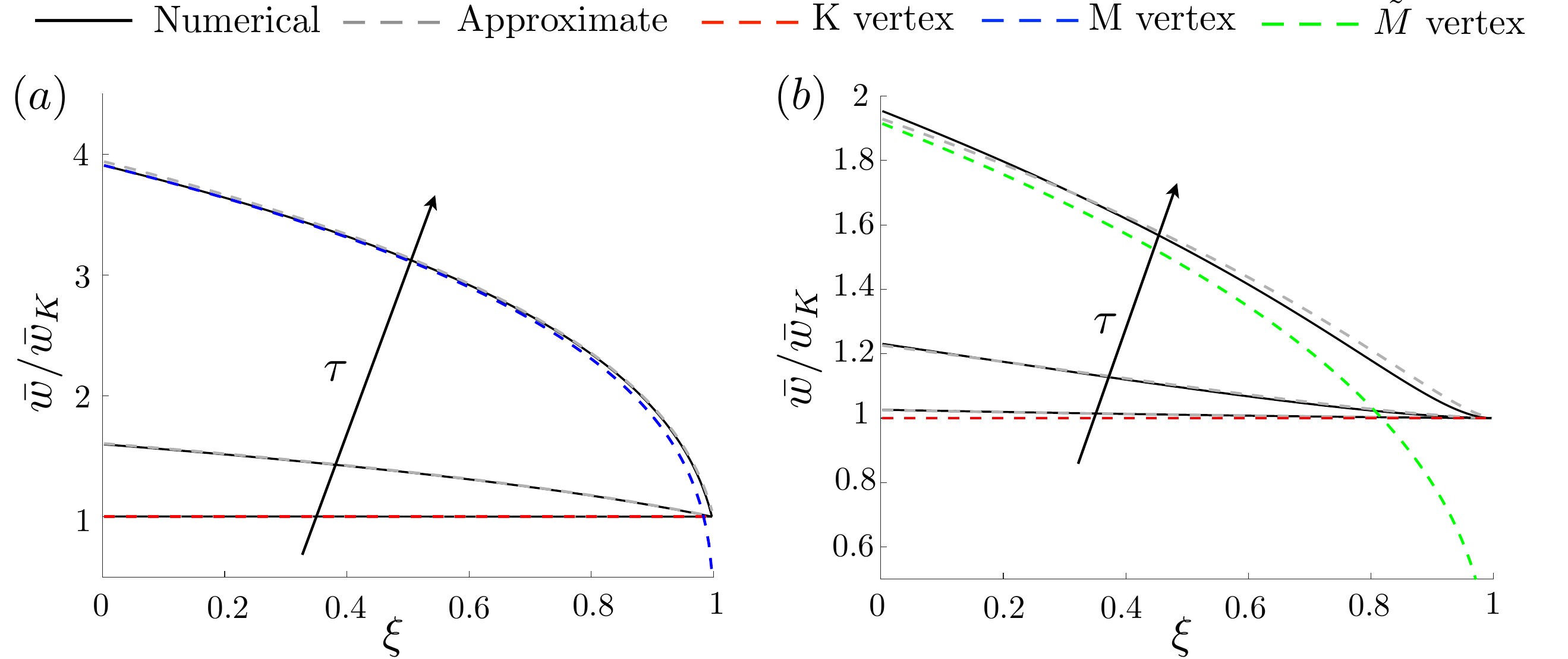}
\caption{Comparison between the numerical, approximate, and vertex solutions in terms of the variation of width versus normalized spatial coordinate $\xi$. Panel $(a)$ corresponds to $\phi=10^{-4}$ and $\tau = \sqrt{4\pi}\{10^{-2},10,10^3\}$. Panel $(b)$ shows the results for $\phi=1$ and $\tau = \sqrt{4\pi}\{0.5,50,5\!\times\!10^3\}$.}
\label{figwvsxi}
\end{figure}
It is also instructive to verify that the solution constructed using the tip asymptote and volume balance accurately approximates the numerical solution. With the reference to~Fig.~\ref{figPKNparamspace}, three values of leak-off parameter are selected $\phi=\{10^{-4},1,10^2\}$. These three values of $\phi$ allow to investigate transitions from the $K$ solution to other three vertices. Fig.~\ref{figlweta} shows the comparison between the numerical solution (solid black lines) and the approximate solution (dashed grey lines) in terms of variation of the normalized fracture length $l/l_K$, width at the wellbore $w/w_K$, and efficiency $\eta$, where the latter is defined as the ratio between the fracture volume and the injected volume. The vertex solutions are indicated by the dashed colored lines. One can observe that the solution indeed transitions from the $K$ limit to either $M$, $\tilde M$, or $\tilde K$ limit. Also, the approximate solution is visually nearly indistinguishable from the one computed numerically. To further compare the two solutions, Fig.~\ref{figwvsxi} plots the normalized width variation along the fracture length for $\phi=10^{-4}$ and $\tau = \sqrt{4\pi}\{10^{-2},10,10^3\}$, and $\phi=1$ and $\tau = \sqrt{4\pi}\{0.5,50,5\!\times\!10^3\}$. Note that the case corresponding to $\phi=10^2$ is trivial since the width is constant for this transition. Results again demonstrate that the difference between the numerical and approximate solutions is small. Thus, confirming that the parametric space, computed using the fast approximate solution, is constructed with a sufficient degree of accuracy. 

\section{Application examples}\label{sectionapplex}
To illustrate the developments, this section presents two examples of how the parametric map can be used in applications.

First, let’s consider a field case with the parameters $H = 20$~m, $K_{Ic} = 1$~MPa$\cdot$m$^{1/2}$, $C'=10^{-6}$~m/s$^{1/2}$, $E'=30$~GPa, $\mu=0.01$~Pa$\cdot$s, $Q=0.01$~m$^3$/s, and $t=1000$~s. With the reference to equations~(\ref{normalization}), (\ref{scales}), and (\ref{phidef}), the dimensionless parameters for this case are $\tau \approx 8 \times 10^4$ and $\phi\approx 4 \times 10^{-4}$. By looking at the parametric map in Fig.~\ref{figPKNparamspace}, it can be concluded that the solution follows the storage viscosity dominated case. Consequently, fracture length can be computed using equation (\ref{solMPKN}) as $l\approx 170$~m.

Second, let’s consider parameters used in~\cite{Dont2016a} to evaluate performance of various numerical algorithms in the toughness regime. The relevant parameters are motivated by laboratory experiments and are given by $H=0.05$~m, $K_{Ic} =1.57$~MPa$\cdot$m, $C'=0$~m/s$^{1/2}$, $E'=3.9$~GPa, $\mu=30.2$~Pa$\cdot$s, $Q=1.7$~mm$^3$/s, and $t=604$~s. According to equations~(\ref{normalization}), (\ref{scales}), and (\ref{phidef}), the dimensionless parameters for this case are $\tau \approx 0.2$ and $\phi=0$. These parameters fall into storage toughness regime in the parametric map shown in Fig.~\ref{figPKNparamspace}. Thus, this confirms that in~\cite{Dont2016a} different algorithms are indeed compared in the toughness regime. Finally, equation~(\ref{solKPKN}) can be used to calculate fracture length as 6.5 cm, which is also in agreement with the numerical results presented in~\cite{Dont2016a}.

\section{Summary}\label{sectionsumm}

The problem of a constant height or PKN hydraulic fracture is investigated. The combined effects of fracture toughness, fluid viscosity, and leak-off are considered. Also, to reduce complexity of the analysis, local elasticity is used in lieu of the more accurate non-local elasticity relation.

The analysis for the problem of a tip region is first outlined. It is shown that, similarly to the classical semi-infinite plane strain hydraulic fracture, there are three limiting or vertex solutions related to domination of either toughness, viscosity, or leak-off. Parametric map of the solutions is obtained, in which zones of applicability of the limiting solutions are outlined. In addition, two approximate solutions for the problem are constructed and their accuracy is evaluated in the whole parametric space.

With respect to the solution for a finite PKN fracture, explicit expressions for the four limiting solutions are first obtained. The latter limits have similar meaning to those for other fracture geometries, such as plane strain and radial, and include storage viscosity, storage toughness, leak-off viscosity, and leak-off toughness limits. The full solution for a finite PKN fracture is computed using two approaches: numerical and fast approximate. The latter utilizes the developed tip asymptotic solution as an approximation for the spatial variation of width for the whole fracture. This fact, combined with the global volume balance, allowed constructing rapid solution for the problem. Accuracy of this approximation is evaluated by comparing its predictions to that from the numerical solution. Equipped with this quick solution, the whole parametric space for the problem is investigated and zones of applicability of the limiting solutions are outlined. One interesting observation is that the global solution evolves from early time storage toughness limit to the late time leak-off viscosity limit. This is different from the behavior for previously analyzed radial and plane strain geometries. The obtained results allow to readily estimate location inside the parametric space for any problem parameters for the constant height fracture, which can be used, for instance, to scale down field data to a laboratory experiment, or to use the closest limiting solution to estimate the fracture geometry.



\end{document}